\documentclass[aps,prd, 10pt, twocolumn,nofootinbib,showpacs]{revtex4-1}
\usepackage{subeqn}
\usepackage[usenames,dvipsnames]{xcolor}

\begin{document}

\title{Overspinning BTZ black holes with test particles and fields}
\author{Koray D\"{u}zta\c{s}}
\email{koray.duztas@emu.edu.tr}
\affiliation{Physics Department, Eastern Mediterranean  University, Famagusta, Northern Cyprus, Mersin 10, Turkey}

\begin{abstract}
It has been claimed that in a test of an asymptotically AdS version of weak cosmic censorship conjecture by attempting to overspin a BTZ black hole with test particles, one finds that it is not possible to spin-up the black hole past its extremal limit. The result of this analysis is restricted  to the case where the initial black hole is extremal. We extend this analysis to find that massive test particles can overspin the black hole, if we start with a nearly extremal black hole, instead. We also consider the interaction of the BTZ black hole with test fields. We show that overspinning  of nearly extremal black holes is possible whether or not there is superradiance for the field. If there is superradiance, overspinning occurs in a narrow range of frequencies bounded below by the super-radiant limit. However, if there is no superradiance for the field, overspinning becomes generic and also applies to extremal black holes. This is in analogy with the Kerr case.
\end{abstract}
\maketitle
\section{Introduction}
The weak form of the cosmic censorship conjecture(wCCC) states that the gravitational collapse of a body always ends up in a black hole rather than a naked singularity. The conjecture was proposed by Penrose in 1969~\cite{ccc} in order to ensure that, 
the singularities which inevitably form as a result of gravitational collapse are hidden behind the event horizons without any access to distant observers. This way, the deterministic nature of general relativity can be preserved since distant observers do not encounter any effects propagating out of singularities.

Over many decades, it has not been possible to establish a concrete proof of wCCC. An alternative thought procedure was developed by Wald~\cite{wald74} to test the stability of event horizons of black holes interacting with test particles or fields. In the first of these thought experiments Wald started with an extremal Kerr-Newman black hole, and showed that particles with enough charge and angular momentum to exceed extremality are not captured by the black hole. Though, some of the similar attempts involving test particles  have been successful to produce naked singularities ~\cite{hubeny,js,richartz,matsas,richartz2}, later backreaction effects were employed to compensate for these violations~\cite{hod,antij,superrad,duztas2}. Thought experiments involving scalar  and electromagnetic fields proved that extremal Kerr black holes cannot be overspun~\cite{semiz,toth1,emccc}, however it is possible to overspin nearly extremal Kerr black holes by using integer spin test fields in a narrow frequency range bounded below by the superradiant limit~\cite{overspin}. In principle, this overspinning can also be invalidated by backreaction effects since the results are analogous to the particle case~\cite{js}. The only generic violation in Kerr case is due to neutrino fields~\cite{duztas,toth,natario}, and it also applies to extremal black holes. Since neutrino fields do not exhibit superradiant behaviour, low frequency modes carrying relatively high angular momentum can also be absorbed. Backreaction effects become negligible as frequency is lowered below the superradiant limit. For that reason violation of wCCC by neutrino fields is generic. (see \cite{paradox} ) 

The three dimensional black hole solution in anti-de Sitter backgrounds which was derived by Ba\~nados, Teitelboim, and Zanelli (BTZ)~\cite{btz}, has many similarities with the Kerr metric. No-hair theorem holds, i.e. a BTZ black hole is fully characterised by its ADM-mass, and angular momentum. Moreover, a rotating BTZ black hole has an inner and outer event horizon like a Kerr black hole. These properties enable similar tests of wCCC to be employed on   charged or neutral BTZ black holes~\cite{btz1,btz2,vitor,gwak2}. Thought experiments involving asymptotically AdS black holes in four and higher dimensions were also constructed~\cite{zhang,rocha,gwak1},  though the notion of distant observers in AdS space-times may not be well defined. 

In this work we first apply a test of wCCC in the case of BTZ black holes 
interacting with massive test particles. This part is motivated by the recent work by Rocha and Cardoso~\cite{vitor}, where they conclude that a BTZ black hole cannot be spun-up past its extremal limit. However, their analysis is restricted to the case where the initial black hole is extremal. We show that the same line of investigation leads to overspinning if we start with a nearly extremal black hole, instead of an extremal one.

We also challenge wCCC by sending in test fields to a BTZ black hole. We follow a similar method to our recent work on overspinning Kerr black holes~\cite{overspin}. Superradiance plays a key role, and the results turn out to be analogous to the Kerr case.

The outline of the paper is as follows: In the introduction we review the general properties of BTZ spacetime and the previous work of Rocha and Cardoso. In section (\ref{particle}) we apply a thought experiment which extends the work of Rocha and Cardoso to include nearly extremal black holes. In section (\ref{field}) we challenge wCCC by sending in test fields to both nearly extremal and extremal black holes. Finally we comment on the results and the possibility to fix the violations of wCCC by backreaction effects.
\subsection{The BTZ solution}
The BTZ black hole is a solution of vacuum Einstein equations in $(2+1)$ dimensions, with a negative cosmological constant $\Lambda=-1/ \ell^2$. The line element is given by
\begin{equation}
ds^2=-N^2dt^2+N^{-2}dr^2+r^2\left(N^{\phi}dt+d\phi\right)^2
\label{btzmetric}
\end{equation}
where $N=N(r)$ and $N^{\phi}=N^{\phi}(r)$ are the usual lapse and shift functions defined by
\begin{equation}
N^2=-M+\frac{r^2}{\ell^2}+\frac{J^2}{4r^2}, \qquad  N^{\phi}=-\frac{J}{2r^2}
\end{equation}
Mass and angular momentum parameters of the BTZ metric are denoted by $M$ and $J$ as usual. Like the Kerr case, the metric has two singularities at 
\begin{equation}
\frac{2r_{\pm}^2}{\ell^2}=M\left(1\pm\sqrt{1-\frac{J^2}{M^2\ell^2}}\right)
\label{rplus}
\end{equation}
(\ref{rplus}) implies that the metric (\ref{btzmetric}) defines a black hole surrounded by an event horizon provided that the mass and the angular momentum parameters satisfy
\begin{equation}
|J|\leq M\ell \label{criterion}
\end{equation}
\subsection{Extremal BTZ black holes and test particles}
One can test the validity of wCCC in the interaction of a BTZ black hole with test particles or fields via (\ref{criterion}). A test particle or a field has negligible effect on the background geometry, it is only expected to slightly perturb the mass and angular momentum parameters. Initially we start with a black hole satisfying (\ref{criterion}), after the interaction with test particles or fields there exists a new configuration with perturbed parameters $J$ and $M$. If the new configuration also satisfies (\ref{criterion}), wCCC remains valid. If the final state violates (\ref{criterion}) wCCC is also violated. The black hole is overspun into a naked singularity, since an event horizon cannot exist if  (\ref{criterion}) is not satisfied.

In that respect, a thought experiment to test the validity of wCCC was recently constructed by Rocha and Cardoso (RC)~\cite{vitor}. In this section we summarise their results. RC consider a black hole with initial parameters $M_0$ and $J_0$. Initially the black hole satisfies (\ref{criterion}), so that $j_0 \equiv J_0/(M_0 \ell)\leq 1$.  They envisage a  particle incident on the black hole from infinity with mass $m_0$, angular momentum $\delta J =m_0 L$, and energy $\delta M= m_0 E$. The geodesic equations lead to
\begin{eqnarray}
\dot{T}     &=& \frac{2ER^2- JL}{2R^2N^2} \\
\dot{R}^2   &=& -\alpha N^2+E^2-L^2/\ell^2+R^{-2}(L^2M- JEL)\label{radialgeod} \\
\dot{\Phi}  &=& \frac{2E J\ell^2-4\ell^2LM+4LR^2}{4R^2\ell^2N^2}
\end{eqnarray}
where $(T(\tau),R(\tau),\Phi(\tau))$ are the coordinates of  the particle, and the dot indicates derivation with respect to proper time $\tau$. ($\alpha=1,0$ for timelike and null geodesics respectively.) Imposing the geodesic to be future directed, $\dot{T} \geq 0$, leads to   $ER^2-(1/2)JL\geq0$ outside the event horizon, via (\ref{radialgeod}). At the horizon, this condition implies that 
\begin{equation}
\frac{L}{E}\leq \frac{2r_+^2}{J}
\label{crit2}
\end{equation}
At the final state the dimensionless spin of the BTZ black hole becomes
\begin{equation}
j\equiv \frac{J_{\rm{fin}}}{M_{\rm{fin}}\ell}=\frac{J_0+m_0\,L}{\ell(M_0+m_0\,E)}
\label{j1}
\end{equation}
In the limit  $m_0 \to 0$, the backreaction on the geometry tends to zero, the system reduces to a black hole and a point test particle. For that reason, RC Taylor expand (\ref{j1}) around $j_0$. 
\begin{equation}
j = j_0-\frac{m_0E}{M_0}\left(j_0-\frac{L}{E\ell}\right)
\label{j2}
\end{equation}
For an extremal black hole, $j_0 =1$ by definition, and (\ref{crit2}) implies and $L/E\leq \ell$. Substituting these in  (\ref{j2}) one finds  that $j \leq 1$. Therefore RC concluded that an extremal BTZ black hole cannot be spun-up past extremality. 
\section{Nearly extremal BTZ black holes and test particles}\label{particle}
An alternative approach to test the validity of wCCC is to start with a nearly extremal black hole instead of an extremal one. This procedure was first employed by Hubeny~\cite{hubeny} who showed that it is possible to overcharge a nearly extremal Reissner-Nordstr\"om black hole as one neglects backreaction effects. A similar analysis was carried out by Jacobson and Sotiriou ~\cite{js} to overspin a nearly extremal Kerr black hole by throwing in test particles. Later backreaction effects were considered for this case to compensate the overspinning of the black hole~\cite{antij}. We also considered integer spin test fields interacting with Kerr black holes~\cite{overspin}, and showed that, though an extremal black hole cannot,   a nearly extremal black hole can actually be overspun by test fields with a frequency slightly above the superradiant limit.  

In this section we carry out a similar analysis to check the validity of wCCC in the interaction of a nearly extremal BTZ black hole with a test particle. We adopt the notation of RC. Let us parametrise a nearly extremal BTZ black hole as
\begin{equation}
j_0=\frac{J}{M \ell}=1-2\epsilon^2
\label{param}
\end{equation}
where $\epsilon \ll 1$. (This parametrization was also used  in \cite{js} and \cite{overspin}.) We work up to second order in $\epsilon$ so $J^2/(M^2 \ell^2)=1-4\epsilon^2$. This leads to
\begin{equation}
\frac{2r_+^{\;2}}{\ell^2}=M(1+2\epsilon)
\label{rplus2}
\end{equation} 
Thus, for a nearly extremal black hole (\ref{crit2}) takes the form
\begin{equation}
\frac{L}{E}\leq \frac{M\ell^2(1+2\epsilon)}{M \ell (1-2\epsilon^2)}=\ell\left(1+2\epsilon +2\epsilon^2 +\rm{O}(\epsilon^3)\right)
\label{crit3}
\end{equation}
For $\epsilon=0$ (\ref{crit3}) reduces to (\ref{crit2}), and $j_0=1-2\epsilon^2 \to 1$. Thus, the results of RC for the extremal BTZ black hole are recovered. For a nearly extremal black hole let us first choose $L/(E \ell)$ arbitrarily close to $(1+2\epsilon +2\epsilon^2)$. Using (\ref{param}), the expression (\ref{j2}) for the final value of the dimensionless spin  takes the form
\begin{equation}
j=1- 2\epsilon^2 + (2\epsilon + 4 \epsilon^2) \frac{m_0 E}{M_0}
\label{jfinal}
\end{equation}
Choosing $\delta M=m_0 E > \epsilon M_0$ leads to $j>1$ so that the BTZ black hole is spun-up past extremality. (With that choice the term $4 \epsilon^2 (m_0 E)/(M_0)$ is neglected) Actually we have to choose $\delta M \sim \epsilon M_0$ to justify the test particle approximation. To summarise the BTZ black hole is spun-up past extremality with two choices: $m_0 E \gtrsim \epsilon M_0$, and $L/(E \ell) \lesssim 1+2\epsilon +2\epsilon^2$.
\section{Overspinning BTZ black holes with test fields}\label{field}
In a recent work we evaluated the validity of wCCC by sending in a packet of waves with dominant frequency $\omega$ and dominant azimuthal wave number $m$ scattering off a Kerr black hole~\cite{overspin}. In this section we carry out a similar analysis for the BTZ black hole. We consider massless fields incident on the black hole from infinity that may be bosonic or fermionic. The BTZ metric admits two Killing vectors $\partial / \partial t$, and $\partial / \partial \phi$ like the Kerr metric, so an incident wave mode has the form 
\begin{equation}
\varphi(r,t,\phi)=e^{-i\omega t}e^{in\phi}f(r)
\end{equation}
(Note that, in BTZ spacetime $\omega$ still denotes frequency, but the azimuthal wave number is referred to with the letter $n$.) We consider a test field which has negligible impact on the background geometry. This field will lead to perturbations $\delta M$ and $\delta J$ in the mass and angular momentum parameters as it scatters off  the BTZ black hole. Our thought experiment involves a monochromatic wave which represents many quanta of energy $\hbar\omega$ and angular momentum $\hbar n$. The perturbations of mass and angular momentum are related by
\begin{equation}
\delta J=(n/\omega)\delta E
\label{beken}
\end{equation}
where $\delta E=\delta M$ for the black hole.

In a thought experiment involving test fields one of the most critical issues is superradiance. The superradiance phenomenon can be roughly defined as the amplification of a wave as it scatters off the black hole. For a BTZ black hole the angular velocity of the horizon is given by $\Omega_{\rm{H}}=J/2r_+^{\;2}$, and the limiting frequency for superradiance is $\omega_{\rm{s}}=n\Omega_{\rm{H}}$. If there is superradiance for a test field incident on the BTZ black hole from infinity, it will occur when the frequency of the field satisfies $\omega<\omega_{\rm{s}}$.

Whether or not superradiance occurs when test fields scatter off BTZ black holes, is an open problem. Naively, from Kerr analogy one would expect it to occur for bosonic test fields, and to be absent for fermionic fields. This expectation was justified by Carlip who argued that superradiance exists for scalar fields on BTZ background~\cite{carlip}. However, Ortiz recently argued that superradiance does not exist for scalar fields on BTZ background, for vanishing boundary conditions at infinity~\cite{ortiz}. 

Let us now consider a nearly extremal BTZ black hole with initial parameters $J_0/(M_0 \ell)=1-2\epsilon^2$, as in (\ref{param}). A test field is incident on the black hole from infinity. For overspinning to occur we require $J_{\rm{fin}}>M_{\rm{fin}}\ell$; that is
\begin{equation}
J_0 + \delta J > (M_0 + \delta E)\ell
\label{fieldcrit1}
\end{equation}
Let us choose $\delta E \sim M_0 \epsilon$ for the incident field, without violating the test field approximation. Using (\ref{beken}) we may rewrite (\ref{fieldcrit1}). (Note that $M_0 \ell-J_0= 2\epsilon^2 M_0 \ell$.)
\begin{equation}
\frac{n}{\omega}M_0 \epsilon > 2\epsilon^2 M_0 \ell + \ell M_0 \epsilon
\label{fieldcrit2}
\end{equation} 
Equation (\ref{fieldcrit2}) implies that the BTZ black hole will be overspun if the frequency of the incoming field is below the critical value:
\begin{equation}
\omega<\frac{n}{\ell (1+2\epsilon)}\equiv \omega_{\rm{c}}
\label{wcritic}
\end{equation}
The critical frequency $\omega_{\rm{c}}$ is slightly larger than the superradiant limit $\omega_{\rm{s}}=n\Omega_{\rm{H}}$ for a nearly extremal black hole. This will become manifest if we write $\omega_{\rm{s}}$ in the form
\begin{equation}
\omega_{\rm{s}}=\left(\frac{J}{M\ell}\right) \frac{n}{\ell (1+2\epsilon)}<\frac{n}{\ell (1+2\epsilon)}=\omega_{\rm{c}}
\end{equation}
where we have used (\ref{rplus2}). Since our initial black hole satisfies $J<M\ell$, $\omega_{\rm{s}}$ is smaller than $\omega_{\rm{c}}$. Therefore if the frequency of the incident field is in the range $\omega_{\rm{s}}<\omega<\omega_{\rm{c}}$ and $\delta E\sim M_0 \epsilon$ for the field, the sub-extremal BTZ black hole will be overspun whether or not there is superradiance for the field. If there is no superradiance for the incoming field the allowed range of frequencies will be extended to $0<\omega<\omega_{\rm{c}}$.

Let us also consider the case where our initial black hole is extremal. By definition $J_0 = M_0 \ell$. The condition for overspinning reduces to
\begin{equation}
\delta J > \ell \delta E
\end{equation}
Let us choose a field with $\delta E \sim M_0 \epsilon^{\prime}$ where $\epsilon^{\prime} \ll 1$. Using (\ref{beken}) one finds that for overspinning to occur,  the frequency of the incoming field should satisfy
\begin{equation}
\omega < \frac{n}{\ell}
\end{equation}
This value coincides with the superradiant limit for an extremal BTZ black hole. Therefore if there is superradiance for the incoming field, the challenging modes will not be absorbed by the black hole so an extremal black hole cannot be overspun. However if there is no superradiance for the field, the modes with $\omega < (n/\ell)$ will also be absorbed, then overspinning will also apply to extremal black holes.
 
\section{Conclusions}
In this work we tested the validity of wCCC in the interactions  of test particles and fields with a  BTZ black hole. For the particle case, Rocha and Cardoso had previously concluded that it was not possible to spin-up a BTZ black hole past extremality. However, their analysis  only applies to extremal black holes. We have shown that the same line of investigation leads to overspinning if we start with a nearly extremal black hole instead. For the case of fields, we found that a nearly extremal black hole can be overspun whether or not there is superradiance for the field, provided that the frequency of the incoming field is in the range $\omega_{\rm{s}}<\omega<\omega_{\rm{c}}$, and $\delta E \sim M_0 \epsilon$ for the field. The critical frequency for extremal black holes coincides with the superradiant limit therefore, if there is superradiance for the field an extremal black hole cannot be overspun.  If there is no superradiance for the field, overspinning also applies to extremal black holes, and the allowed range of frequencies is extended to $0<\omega<\omega_{\rm{c}}$. 

Backreaction effects were neglected in this analysis. In principle, backreaction effects can be employed to compensate for the overspinning of BTZ black holes by test particles, and test fields that exhibit superradiant scattering. However, if there is no superradiance for the field, the violation of wCCC becomes generic and also applies to extremal black holes. The reason for that is the fact that the ratio $\delta J /\delta E$ will grow without bound and backreaction effects will become negligible as the frequency $\omega$ is lowered below the superradiant limit. This is in analogy with the Kerr case, where the overspinning due to neutrino fields is generic. 

Superradiance can turn into an instability with a mirror mechanism that reflects the radiation backwards, causing it to be amplified for multiple times.  However, as far as the scattering of test fields is concerned, superradiance acts as a cosmic censor as it prevents the absorption of dangerous modes with low energy and high angular momentum. In other words it works in favour of the stability of the event horizon, therefore the stability of the black hole. In the absence of superradiance one can always find a frequency sufficiently low, so that the ratio $\delta J /\delta E$ is sufficiently high to overspin an extremal or a nearly extremal black hole. From that point of view superradiance is necessary for the stability of the black hole.

\end{document}